\documentclass[floatfix]{revtex4}
\usepackage[dvips]{graphicx}

\begin{document}

\begin{flushright} 
physics/0401007 \\
January 2004
\end{flushright}

\title{\Large The CAVES Project - Exploring Virtual Data Concepts for Data Analysis\\
\vspace{0.75cm}
\raggedleft \normalsize To Dad}

%

\author{D.~Bourilkov}
\affiliation{University of Florida, Gainesville, FL 32611, USA}

\begin{abstract}
The Collaborative Analysis Versioning Environment System ({\tt CAVES}) project
concentrates on the interactions between users performing data and/or
computing intensive analyses on large data sets, as encountered in many
contemporary scientific disciplines. In modern science increasingly
larger groups of researchers collaborate on a given topic over
extended periods of time. The logging and sharing of knowledge about how
analyses are performed or how results are obtained is important
throughout the lifetime of a project. Here is where virtual data
concepts play a major role. The ability to seamlessly log, exchange
and reproduce results and the methods, algorithms and computer programs
used in obtaining them enhances in a qualitative way the level of
collaboration in a group or between groups in larger organizations.
It makes it easier for newcomers to start being productive almost from
day one of their involvement or for referees to audit a result and
gain easy access to all the relevant details. Also when scientists move
on to new endeavors they can leave their expertise in a form easily
utilizable by their colleagues. The same is true for archiving the
knowledge accumulated in a project for reuse in future undertakings.

      The {\tt CAVES} project takes a pragmatic approach in assessing the
needs of a community of scientists by building series of prototypes
with increasing sophistication. In extending the functionality of
existing data analysis packages with virtual data capabilities these
prototypes provide an easy and habitual entry point for researchers to
explore virtual data concepts in real life applications and to provide
valuable feedback for refining the system design. The architecture
is modular based on Web, Grid and other services which can be plugged in
as desired. As a proof of principle we build a first system by extending the
very popular data analysis framework {\tt ROOT}, widely used in high energy
physics and other fields, making it virtual data enabled.

\end{abstract}

\maketitle



\section{INTRODUCTION}

Most data in contemporary science are the product of increasingly complex
computations and procedures applied on the fast increasing flows of raw
information coming from more and more sophisticated measurement devices
(the ``measurements''), or from growingly detailed numeric simulations - 
e.g. pattern recognition, calibration, selection,
data mining, noise reduction, filtering, estimation of parameters etc.
High energy physics and many other sciences are increasingly CPU and data
intensive. In fact, many new problems can only be addressed at the high
data volume frontier. In this context, not only data analysis transformations,
but also the detailed log of how those transformations were applied,
become a vital intellectual resource of the scientific community.
The collaborative processes of these ever-larger groups require
new approaches and tools enabling the efficient sharing of knowledge and
data across a geographically distributed and diverse environment.

Here is where the concept of virtual data is bound to play a central role
in the scientific analysis process. We will explore this concept using as
a case study the coming generation of high energy physics (HEP) experiments
at the Large Hadron Collider (LHC), under construction at the European
Laboratory for Particle Physics CERN close to Geneva, Switzerland. This
choice is motivated by the unprecedented amount of data (from petabytes
to exabytes) and the scale of the collaborations that will analyze it (four
worldwide collaborations, the biggest two with more than two thousand
scientists each). At the same time, the problems to be solved
are general and will promote scientific discoveries in different disciplines,
enhance business processes and improve security. The challenge facing HEP
is a major driving force for new developments in computing, e.g. the
Grid~\cite{gridblueprint,gridanatomy}.
The computing landscape today is marked by the rise of
Grid~\cite{gridphysiology} and Web services~\cite{w3cwebserv}
and service oriented architectures (SOA)~\cite{soa}.
An event-driven SOA is well suited
for data analysis and very adaptable to evolution and change over time. In
our project we will explore and adopt service oriented solutions as they
mature and provide the performance needed to meet mission-critical requirements.

This paper is organized as follows:
in the next section we introduce the concept of
virtual data,
than we discuss the issues arising when dealing with
data equivalence,
describe how
data analysis is done in HEP,
digress with
a metaphor,
elucidate the ideas driving the
{\tt CAVES} project,
continue with a detailed treatment of the
{\tt CAVES} architecture,
sketch the
first implementation,
discuss the 
relationship with other Grid projects,
and conclude with an
outlook.

\section{VIRTUAL DATA}

The scientific analysis process demands the precise tracking of
how data products are to be derived, in order
to be able to create and/or recreate them on demand.
In this context virtual data are data products with a well defined method
of production or reproduction. The concept of ``virtuality'' with respect to
existence means that we can define data products that may be produced in the
future, as well as record the ``history'' of products that exist now
or have existed at some point in the past.

The virtual data paradigm logs data provenance by tracking
how new data is derived from transformations on other
data~\cite{chimera,chimera2}.
{\em Data provenance} is the exact history of any existing (or virtual)
data product.
Often the data products are large datasets, and the management of dataset
transformations is critical to the scientific analysis process.

We need a ``virtual data management'' tool that can ``re-materialize'' data
products that were deleted, generate data products that were defined but
never created, regenerate data when data dependencies or algorithms change,
and/or create replicas at remote locations when recreation is more efficient
than data transfer.

From the scientist's point of view, data trackability and result auditability
are crucial, as the reproducibility of results is fundamental to the nature of
science. To support this need we require and envision something like a
``virtual logbook'' that provides the following capabilities:
\begin{itemize}
\item easy sharing of tools and data to facilitate collaboration -
all data comes complete with a ``recipe'' on how to produce or reproduce it;
\item individuals can discover in a fast and well defined way other scientists'
work and build from it;
\item different teams can work in a modular, semi-autonomous fashion; they
can reuse previous data/code/results or entire analysis chains;
\item on a higher level, systems can be designed for workflow management and
performance optimization, including the tedious processes of staging in data
from a remote site or recreating it locally on demand (transparency with
respect to location and existence of the data).
\end{itemize}

\section{DATA EQUIVALENCE}

If we delete or accidentally loose a piece of data, having a log of how
it came into existence will come in handy. Immediately the question arises:
is the ``new'' chunk of data after reproduction identical to the ``old'' one?
There are two extreme answers to this question:

\begin{itemize}
\item the two pieces of data are identical bitwise - we are done;
\item not only are the two pieces of data not identical bitwise,
but they contain different information from the viewpoint of the application
using them.
\end{itemize}
The second point needs discussion: clearly two chunks of data can be
``identical enough'' for some types of applications and different for
other types. Each application has to define some ``distance'' measure
between chunks of data and specify some ``minimal'' distance between
chunks below which the pieces are considered identical. In this language
bitwise sameness would correspond to zero distance.

    Let us illustrate this with two examples. In an ideal world, if we
generate events with the Monte Carlo method, starting from the same seeds
and using portable random number generators, we should get the same
sequence of events everywhere. Or if we do Monte Carlo integration, we
should get exactly the same result. In practice, due to floating point
rounding errors, even on systems with processors with the same word length
simulations tend to go down different branches, diverging pretty soon.
So the results are not guaranteed to be identical bitwise. Usually this is
not a problem: two Monte Carlo integrations within the statistical
uncertainty are certainly acceptable. And even two different sequences
of events, when their attributes are statistically equivalent (e.g.
histograms of all variables, correlations etc.), are good enough for many
practical purposes. There are exceptions though: if our code crashes at event
10583, we would like to be able to reproduce it bitwise. One way to proceed
in such a situation is to store the initial random seeds for {\em each} event
along with the how-to (i.e. the algorithm and code for producing events). Then
any single divergence will affect at most one event.

    The second example is analysis of real data. If we are interested in
statistical distributions (histograms, scatter plots, pie charts etc.),
a ``weak'' equivalence in the statistical sense can be enough. If we
are selecting e.g. rare events in a search for new particles, we would
like to isolate the same sample each time we run a particular selection
on the same input (``strong'' equivalence). One way to proceed here is to keep
the list of selected events along with the how-to of the selection for future
verifications. As long as the input sample is available, we will have
reproducibility of the selection even if portability is not guaranteed.

    To sum it up - each application has to define criteria establishing
the equivalence of data for its domain. Good choice of metadata about
a chunk of data (e.g. a dataset) can be very useful later when trying to
decide if your reproduction is good enough. For instance, if we kept the
moments like mean value and standard deviation with their statistical
uncertainties from a distribution with millions of events, it will help
in determining if our replica is statistically equivalent later.

    Last but not least, an important aspect in recording the data provenance
is the level of detail. The result of the execution of the same algorithm
with the same input in today's complex software world may depend on
environment variables, linked libraries containing different versions of
supporting applications, different compilers or levels of optimization etc.
When these factors are important, they have to be included in the data
provenance log for future use.

\section{DATA ANALYSIS IN HEP}

The high energy physics field is sociologically very interesting. The
experimental collaborations have grown from being counted on the fingers of
one or two hands in the sixties to around five hundred in the nineties
and two thousand today. Even in theoretical physics collaborations are
growing with time. That explains why the field was always in the forefront
of developing and/or adopting early new collaborative tools, the best known
example being of course the invention of the World Wide Web at CERN. At
present, the LHC experiments are heavily involved in Grid efforts, continuing
the tradition, see e.g.~\cite{griphyn,ppdg,eudatagrid,ivdgl,lcg,egee}.

After a high energy physics detector is triggered, the information from the
different systems is read and ultimately recorded (possibly after cleaning,
filtering and initial reconstruction) to mass storage. The high intensity
of the LHC beams usually results in more than one interaction taking
place simultaneously, so a trigger records the combined response to all
particles traversing the detector in the time window when the system is open.
The first stages in the data processing are well defined and usually tightly
controlled by the teams responsible for reconstruction, calibration, alignment,
``official'' simulation etc. The application of virtual data concepts in
this area is discussed e.g. in~\cite{cmsprod}.

Here we are interested in the later stages of data processing
and analysis, when various teams and individual scientists look at the data
from many different angles - refining algorithms, updating calibrations or
trying out new approaches, selecting and analyzing a particular data set,
estimating parameters etc., and ultimately producing and publishing physics
results. Even in today's large collaborations this is a decentralized,
"chaotic" activity, and is expected to grow substantially in complexity and
scale for the LHC experiments. Decentralization does not mean lack of
organization - on the contrary, this will be one of the keys for building
successful structures, both from the social and technical points of view.
Clearly flexible enough systems, able to accommodate a large user base, and
use cases not all of which can be foreseen in advance, are needed. Many users
should be able to work and share their results in parallel, without stepping
on each other's toes. Here we explore the benefits that a virtual data
system can bring in this vast and dynamic field.

\begin{figure*}[htb]
\centering
\resizebox{0.97\textwidth}{0.37\textheight}{\includegraphics{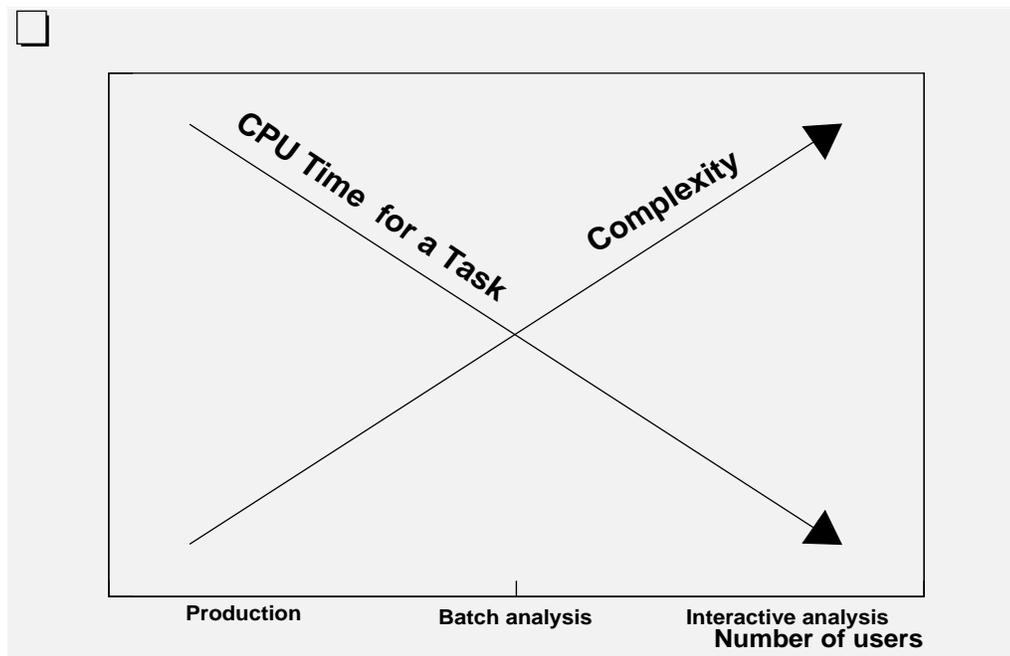}}
\caption{From production through batch to interactive analysis:
the complexity grows fast with
the number of users, shown on the horizontal axis in arbitrary units,
while the average wall and CPU time to complete a typical task goes down.}
\label{useranal}
\end{figure*}

Moving from production to analysis, the complexity grows fast with the number
of users while the average wall and CPU time to complete a typical task goes
down, as illustrated in Figure~\ref{useranal}.
An intermediate phase are large analysis tasks which require
batch mode. The ultimate challenge comes from interactive analyses, where
users change their minds often and need fast response times to stay
productive. The latency of the underlying systems at this stage is
critical. There should be no single point of failure and the system should
re-route requests automatically to the next available service. The redundancy
should be accompanied by efficient synchronization, so that new results
are published fast and made accessible for all interested parties regardless of
their geographical location.

An important feature of analysis systems is the ability to build scripts
and/or executables ``on the fly'', including user supplied code and parameters.
On the contrary, production systems often rely on pre-build applications,
distributed in a centralized way from ``officially controlled'' repositories.
The user should be in position to modify the inputs on her/his desk(lap)top
and request a derived data product, possibly linking with preinstalled 
libraries on the execution sites. A grid-type system can store large
volumes of data at geographically remote locations and provide the necessary
computing power for larger tasks. The results are returned to the user or
stored and published from the remote site(s). An example of this vision
is presented in~\cite{Arbree:2003wh}.
At each stage in the analysis interesting events can be visualized and plots
for all quantities of interest can be produced.

What will a virtual data system bring to this picture? I have in my office
a large collection of paper folders for different analyses performed working
on various tasks, stored in varied ways. They match with codes stored and
archived on different systems. So when a colleague comes in and asks me how
I obtained that plot three months ago, I have to sift for some time (depending
on my organization) through my folders, make a photocopy, then find the
corresponding code, make sure it is the ``historically'' right version, and
that I wrote down which version of the pattern recognition was used at the
time etc. Or if I go to my colleague, she will go through
similar steps, but her organization will be different and we will possibly
exchange information in a different format. Or if one of us is on leave things
will slow down. So we are recording data provenance, but manually, very
often incomplete, and not easily accessible. And clearly this scales poorly
for larger and geographically distributed collaborating groups.

In a ``virtual logbook'' all steps of an analysis, even the blind alleys,
can be recorded and retrieved automatically.
Let us assume that a new member joins the group. Even without bugging  one's
colleagues too often, it will be quite easy to discover exactly what
has been done so far for a particular analysis branch, to validate how it was
done, and to refine the analysis.
A scientist wanting to dig deeper can add a new derived data branch
by, for example, applying a more sophisticated selection, and continuing to
investigate down the new road.
Of course, the results of the group can be shared easily with other teams
and individuals in the collaboration, working on similar topics, providing
or re-using better algorithms etc. The starting of new subjects will
profit from the availability of the accumulated experience.
At publication time it will be much easier to perform an accurate audit of the
results, and to work with internal referees who may require details of the
analysis or additional checks.

\section{A METAPHOR}

At the beginning of a new project, a suitable metaphor can be
helpful. As we would like to make our ``virtual data logbooks''
persistent, distributed and secure, the following analogy came quite
naturally:
\begin{itemize}
\item A cave is a secure place to store stuff.
\item Usually you need a key to enter.
\item Stuff can be retrieved when needed (and if the temperature
is kept constant, usually in good shape).
\item Small caves can be private, larger ones are usually owned by
cooperatives.
\item When a cave is full, a new one is build.
\item To get something, one starts at the local caves and, if needed,
widens the search ...
\end{itemize}
We can go on, but, as we will see in a moment, these are striking
similarities with the goals of our project, so {\tt CAVES} seemed a
peculiarly apt name.
The use of metaphors is inspired from the adoption of extreme
programming techniques~\cite{beck} in our project.
For their relationship to the programming style in HEP see~\cite{carmin}.

\section{CAVES PROJECT}

The Collaborative Analysis Versioning Environment System ({\tt CAVES})
project concentrates on the interactions between users performing 
data and/or computing intensive analyses on large data sets, 
as encountered in many contemporary scientific disciplines. 
In modern science increasingly larger groups of researchers 
collaborate on a given topic over extended periods of time. 
The logging and sharing of knowledge about how analyses are 
performed or how results are obtained is important throughout 
the lifetime of a project. Here is where virtual data concepts 
play a major role. The ability to seamlessly log, exchange 
and reproduce results and the methods, algorithms and computer 
programs used in obtaining them enhances in a qualitative 
way the level of collaboration in a group or between groups 
in larger organizations.

It makes it easier for newcomers to start being productive 
almost from day one of their involvement or for referees to 
audit a result and gain easy access to all the relevant details. 
Also when scientists move on to new endeavors they can leave 
their expertise in a form easily utilizable by their colleagues. 
The same is true for archiving the knowledge accumulated in 
a project for reuse in future undertakings.

The {\tt CAVES} project takes a pragmatic approach in assessing the 
needs of a community of scientists by building series of prototypes 
with increasing sophistication. Our goal is to stay close to the
end users and listen carefully to their needs at all stages of an
analysis task. In this way we can develop an architecture able to satisfy
the varied requirements of a diverse group of researchers.

Our main line of development draws on the needs of, but is not limited
to, high energy physics experiments, especially the
CMS collaboration~\cite{cms,cmssoft,cmsweb}, planning to begin data taking
in 2007 at the Large Hadron Collider. The CMS experiment will 
produce large amounts of simulated and real data, reaching 
tens and hundreds of petabytes. The analysis of datasets of 
this size, with its distributed nature, by a large community 
of users is a very challenging task and one of the strongest 
driving forces for Grid computing. The {\tt CAVES} project explores
and develops these emerging technologies to facilitate the 
analysis of real and simulated data. We start by analyzing 
the simulated data from the data challenges of the CMS experiment,
which will grow in scale and complexity approaching the situation 
when real data will start to flow.

In extending the functionality of existing data analysis packages 
with virtual data capabilities, we build functioning analysis suites, 
providing an easy and habitual entry point for researchers 
to explore virtual data concepts in real life applications, 
and hence give valuable feedback about their needs, helping 
to guide the most useful directions for refining the system design.
By just adding capabilities in a plug-in style we facilitate the
acceptance and ease of use, and thus hope to attract a critical mass
of users from different fields in a short time. There is no need
to learn yet another programming language, and our goal is simplicity
of design, keeping the number of commands and their parameters to
the bare minimum needed for rich and useful functionality.

The architecture is modular based on Web, Grid and other services 
which can be plugged in as desired. In addition to working 
in ways considered standard today the scientists are able 
to log or checkpoint their work throughout the lifetime of 
an analysis task. We envisage the ability to create private 
``checkpoints'' which can be stored on a local machine 
and/or on a secure remote server. When an user wants to share 
some work, he can store the relevant know-how on the group 
servers accessible to the members of a group working on a given 
task. This could be a geographically distributed virtual organization. 
In the case of collaboration between groups or with internal 
or external referees portions of this know-how can be made 
accessible to authorized users, or a shared system of servers 
can be created as needed. The provenance of results can be 
recorded at different levels of detail as decided by the users 
and augmented by annotations. Along with the knowledge of 
how analyses are performed, selected results and their annotations 
can be stored in the same system. They can be browsed by the 
members of a group, thus enhancing the analysis experience 
both for experts and newcomers. When desirable, information 
from different phases of an analysis can easily be shared 
with other groups or peers.

We stressed already the value of complete logs. In the heat of an
active analysis session, when there is no time or need to be pedantical,
users may see merit in storing sometimes also partial logs, a classical
example being a program with hidden dependencies, e.g. the calling of
a program or reading of a file within a program, not exposed externally.
In this case, the data product is not reproducible, but at least the
log will point what is missing. Or the users may even store a non-functional
sequence of actions in the debugging phase for additional work later,
even without producing a virtual data product. Our system
should be able to support partial logging, provided that the users are
aware of the limitations and risks of this approach.

An important point is how groups will structure their analyses. 
Each virtual data product needs an unique identifier, which may be
provided by the users or appended automatically by the system
with e.g. project id, user id and date to render it unique.
For smaller tasks, all identifiers and logs can be located in a
single place, like a big barrel in a cave. Then the group will
benefit from adopting a policy for meaningful selection of identifiers,
making subsequent browsing and finding of information easy.
For larger projects the virtual data space can be structured
in chunks corresponding to subtasks, like many barrels in a large
cave. Then at the beginning of a session the user will select
the barrel to be opened for that session. When needed, information
from related (linked) barrels can be retrieved. In principal 
there are no restrictions on how deep the hierarchy can be, only
the practical needs will determine it.

We base our first functional system on popular and well established data
analysis frameworks and programming tools, making them virtual data enabled. 
In the course of our project we will leverage best-of-breed existing
technologies (e.g. databases, code management systems, Web services),
as well as the developments in forward-looking Grid enabled projects,
e.g. the virtual data system {\tt CHIMERA}~\cite{chimera,chimera2},
the {\tt CLARENS} server~\cite{clarens} for secure remote dataset access,
the {\tt Condor}~\cite{condor},
{\tt PEGASUS}~\cite{pegasus} and {\tt SPHINX}~\cite{sphinx}
schedulers for executing tasks in a Grid environment.
Further down the road we envisage building distributed systems capable 
of analyzing the datasets used in the CMS collaboration at 
all stages of data analysis, starting from Monte Carlo generation 
and simulation of events through reconstruction and selection 
all the way to producing results for publication.
We plan to use the Grid test bed of the Grid Physics Network 
(GriPhyN) project~\cite{griphyn} for Grid enabling 
the collaborative services. The GriPhyN Project is developing 
Grid technologies for scientific and engineering projects 
that will collect and analyze distributed, petabyte-scale 
datasets. GriPhyN research will enable the development of 
Petascale Virtual Data Grids (PVDGs) through its Virtual Data 
Toolkit (VDT~\cite{vdt}).

\section{CAVES ARCHITECTURE}

The {\tt CAVES} system can be used as a building block for a collaborative
analysis environment, providing ``virtual data logbook'' capabilities and
the ability to explore the metadata associated with different data products.

Our first functioning system extends the very popular object-oriented data
analysis framework {\tt ROOT}~\cite{root}, widely used in high energy physics
and other fields, making it virtual data enabled. The {\tt ROOT} framework
provides a rich set of data analysis tools and excellent graphical capabilities,
able to produce publication-ready pictures. It is easy to execute user
code, written in C++, and to extend the framework in a plug-in style. New
systems can be developed by subclassing the existing {\tt ROOT} classes. And
the {\tt CINT}~\cite{cint} interpreter runs the user code ``on-the-fly'',
facilitating fast development and prototyping. All this is very helpful in the
early phases of a new project. In addition, {\tt ROOT} has a large and lively
user base, so we plan to release early and often and to have a development
driven largely by user feedback. Last but not least, {\tt ROOT} is easy
to install and very portable. Versions for many flavors of LINUX and UNIX
and for Windows are available.

We leverage a well established source code management system - the
Concurrent Versions System {\tt CVS}~\cite{cvs}.
It is well suited to provide version control for a rapid development by
a large team and to store, by the mechanism of tagging releases, many versions
so that they can be extracted in exactly the same form even if modified,
added or deleted since that time. The {\tt CVS} tags assume the role of
unique identifiers for virtual data products.
{\tt CVS} can keep track of the contributions of different users. The locking
mechanism makes it possible for two or more people to modify a file at
the same time, important for a team of people working on large projects.
The system has useful self-documenting capabilities. Besides the traditional
command line interface several products provide Web frontends which can be
used when implementing Web services.
All these features of {\tt CVS} make it a good match for our system.
Nowadays {\tt CVS} is already installed by default on most UNIX and LINUX
systems and a Windows port is available. In this way, our system can be used
both from UNIX and Windows clients, making it easy for users at all levels to
reproduce results.

A key aspect of the project is the distributed nature of the input data,
the analysis process and the user base. This has to be addressed 
from the earliest stages. Our system should be fully functional
both in local and remote modes, provided that the necessary repositories
are operational and the datasets available. This allows the users
to work on their laptops (maybe handhelds tomorrow) even without a
network connection, or just to store intermediate steps in the course of an
active analysis session for their private consumption, only publishing
a sufficiently polished result. This design has the additional benefit
of utilizing efficiently the local CPU and storage resources of the users,
reducing the load on the distributed services (e.g. Grid) system.
The users will have the ability to replicate, move, archive and delete
data provenance logs. Gaining experience in running the system will
help to strike the right balance between local and remote usage.
More details about the distribution of services is given in the next
section.

\begin{figure*}[htb]
\centering
\resizebox{0.97\textwidth}{0.48\textheight}{\includegraphics{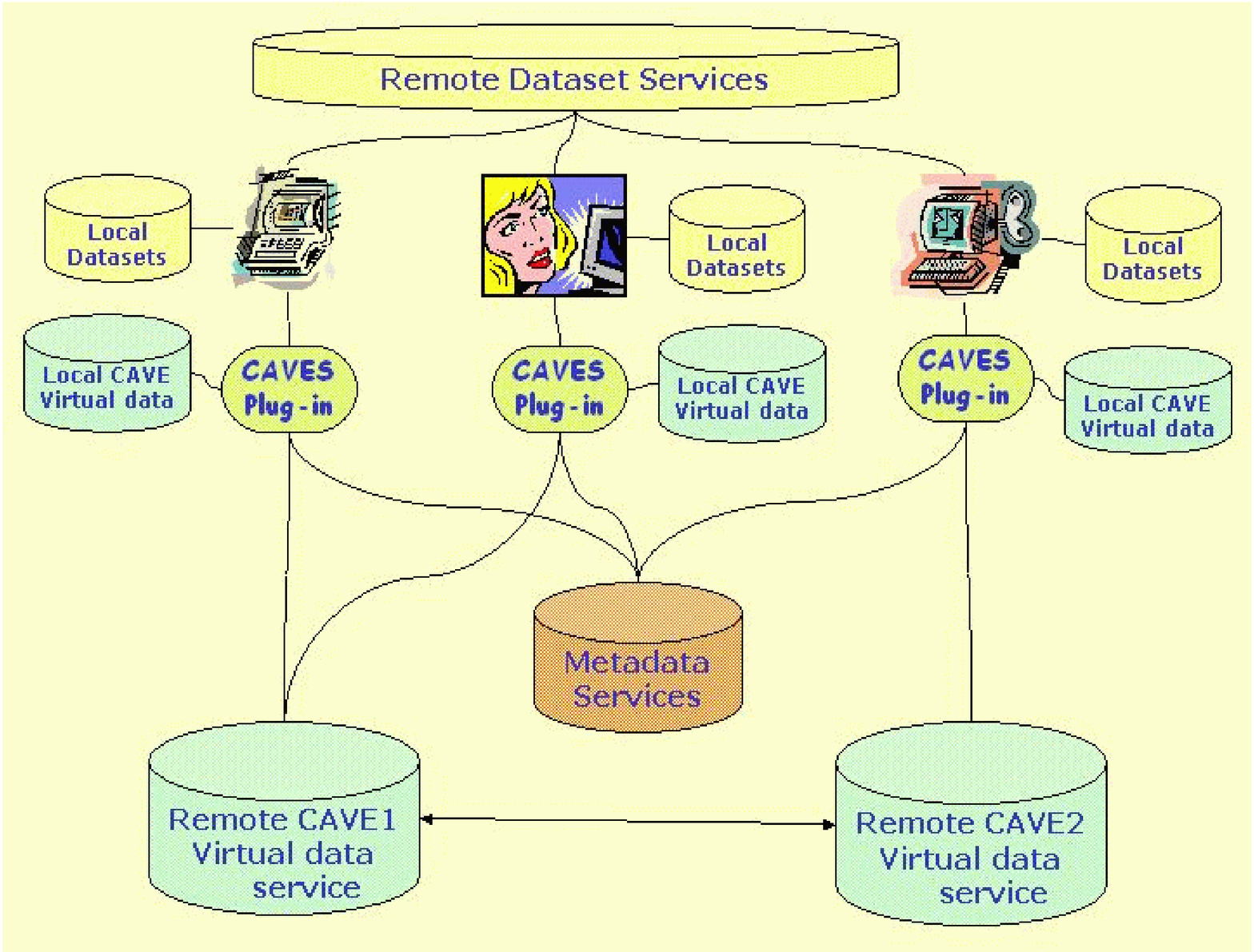}}
\caption{The distributed and scalable {\tt CAVES} architecture.
Three different users performing analyses are shown, but there is no upper
limit on the number or type of users (LINUX, WINDOWS etc.).
The users can work both in local and remote mode. A {\tt CAVES} plug-in makes
them virtual data enabled. The users can log the complete history of their
analyses between checkpoints in local caves (on their machines), or in
distributed remote virtual data logbooks. Moreover, they can store annotations
and even selected results. The line between the remote caves shows that caves
can be mirrored, synchronized etc. depending on the needs of collaborating
groups. In addition, more centralized metadata services can store the
annotations and condensed logs for fast search and browsing, reducing the load
on the virtual data services. The datasets can be stored remotely and (part of
them) locally. In this way the users can alternate between local and remote
mode. The knowledge accumulated during the lifetime of an analysis can be
automatically logged, shared and reproduced on demand.}
\label{cavesarch}
\end{figure*}

The architecture of {\tt CAVES} builds upon the concept of sandbox 
programming. By sandbox programming we mean users work
on per session basis, creating a new sandbox for a given session. 
All the changes or modifications and work the user does in a
session between checkpoints is logged into a temporary logfile, which can 
be checked in the {\tt CVS} repository with a unique tag.
The system checks if the user executed external programs
(in the {\tt ROOT} case these are typically C++ programs) and
logs them automatically with the same tag. Here an interesting
point arises: a possible scenario is that a user runs the same
program many times, just varying the inputs. In this case
{\tt CVS} will do the right thing: store the program only {\em once},
avoiding duplication of code, and tagging it many times with
different tags, reflecting the fact that the program was executed
several times to produce distinct data products.
Or the user can choose during the same session to browse through the tags
of other users to see what work was done, and select the 
log/session of interest by extracting the peers log with the tag used
to log the corresponding session activities. Here two modes of operation
are possible: the user may want to reproduce a result by extracting
and executing the commands and programs associated with a selected tag,
or just extract the history of a given data product in order to inspect
it, possibly modify the code or the inputs and produce new results.

We also have the concept that users can log annotations 
or results in the repository along with the data provenance, storing
useful metadata about a data product for future use (see the discussion in
the data equivalence section). It is possible to record the metadata in
relational databases too, e.g. in popular open source products
like MySQL~\cite{MySQL},
so that another user first queries the database to retrieve the annotations or 
condensed logs of what other users have done already. This approach will
ensure scalability for large groups of researchers accumulating large
repositories, and will reduce the load on the {\tt CVS} servers, improving the
latency of the overall system. In this case the searching of a database is
expected to be faster than the direct search for a given tag among a large
number of stored tags. This will be investigated by building functional systems
and monitoring their performance. The additional burden of synchronizing the
information between the databases and the repositories is worthwhile only if we
can improve the overall performance and scalability of the system. A further
enhancement can come from retrieving first the metadata, and only if the user
is interested, the complete log about a particular data product.

The architecture is shown in graphical form in Figure~\ref{cavesarch}.

Let us discuss now some possible scenarios, 
which can take place in our architecture.

{\bf Case1: Simple} \\
User 1 : Does some analysis and produces a result with tag 
{\em {\bf projectX-stepY-user1}}. \\
User 2: Browses all current tags in the repository and fetches 
the session stored with tag {\em {\bf projectX-stepY-user1}}.\\

{\bf Case2: Complex} \\
User 1 : Does some analysis and produces a result with tag 
{\em {\bf projectX-stepY-user1}}. \\
User 2: Browses all current tags in the repository and fetches 
the session stored with tag {\em {\bf projectX-stepY-user1}}. \\
User 2: Does some modifications in the code files, which 
were obtained from the session of user1, runs again and stores the changes
along with the logfile with a new tag
\mbox{{\em {\bf projectX-stepY-modcode-user2}}}. \\
User 1: Browses the repository and discovers that the previous 
session was used and contains modified or new code files, so decides 
to extract that session using the new tag
\mbox{{\em {\bf projectX-stepY-modcode-user2}}} and possibly reuse it to
produce the next step and so on. \\
This scenario can be extended to include an arbitrary number of steps and users
in a working group or groups in a collaboration.

Based on our work so far, the following set of commands emerges as useful:
\begin{enumerate}
\item {\em Session commands:}
\begin{itemize}
  \item {\bf open $\rm <session>$} : authentication and authorization,
        connection, selection of {\tt CVS} services, local or remote mode, 
        the barrel to be opened etc.
  \item {\bf close $\rm <session>$} : save opened sessions, clean-up etc.
\end{itemize}
\item {\em During analysis:}
\begin{itemize}
  \item {\bf help $\rm <command>$} : get help for a command or list of commands
  \item {\bf browse $\rm <tag>$} : browse all tags in (a part of) a repository,
        subsets of tags beginning or containing a string etc; possibly browse
        the metadata about a specific virtual data product e.g. by clicking
        on a selected tag from a list displayed in a graphical user interface
  \item {\bf startlog} : define the starting checkpoint for a log (part of a
        session between user-defined points), which will be closed by a
        {\bf log} command
  \item {\bf log $\rm <tag>$} : log (part of) a session between user-defined
        checkpoints together with all programs executed in the session;
        this may be a complete or optionally a partial log with user-defined
        level of detail
  \item {\bf annotate $\rm <tag>$} : store user-supplied notes (metadata) about
        the work being done, preferably in a concise and meaningful manner; 
        optionally, selected results can be stored along with the annotations
        e.g. a summary plot, subject of course to space considerations; this
        can be a separate command or part of the {\bf log} command e.g.
        the user may select to be prompted to provide annotations when
        logging a tag
  \item {\bf inspect $\rm <tag>\ \ <brief|complete>$} : get a condense
        (annotations plus user commands, in a sense something like header
        files), or the complete log for a tag including the programs executed,
        but do not reproduce the data product; useful for reusing analysis work
  \item {\bf extract $\rm <tag>$} : in addition to {\bf inspect}, reproduce the
        virtual data product
\end{itemize}
\item {\em Administrative tasks:}
\begin{itemize}
  \item {\bf copy $\rm <tag>$ $\rm <from>$ $\rm <to>$} : clone a log to a new
        repository
  \item {\bf move $\rm <tag>$ $\rm <from>$ $\rm <to>$} : as you expect
  \item {\bf delete $\rm <tag>$ $\rm <from>$} : remove a log; {\tt CVS} has the
        nice feature of storing such files in the Attic, so it is similar to
        moving a file in the Trash can without emptying it
  \item {\bf archive $\rm <tag>$ $\rm <to>$} : store in an archive (e.g. a mass
        storage system)
  \item {\bf retrieve $\rm <tag>$ $\rm <from>$} : retrieve from an archive (for
        whole repositories normal {\tt CVS} techniques can be used).
\end{itemize}
\end{enumerate}

Our first release is based on the minimal scope of commands providing
interesting functionality. Future developments will be guided by the value
users put on different options. As we want a fast release cycle this limits
the number of features introduced and tested in any new version. The command
set is extensible and new commands may be introduced as needed.

\section{FIRST IMPLEMENTATION}

In this section we sketch the process of building the first {\tt CAVES} release
as prototype of a real analysis system, and examine some of the issues that
arise. The first prototype has been demonstrated at the Supercomputing
conference in Phoenix, Arizona, in November 2003, and the first release was
made public on December 12, 2003. More details can be found
in~\cite{cavestalk01}, and an in-depth technical description about the virtual
data enabled {\tt ROOT} client and the remote services is in
preparation~\cite{caves01}.

We limit the scope to the most basic commands described in the previous
section:
\begin{itemize}
  \item {\bf open $\rm <session>$} : sets the {\tt CVS} service (default or
        user choice)
  \item {\bf help}: short help about the commands below
  \item {\bf browse $\rm <tag>$} : browse tags in the repository
  \item {\bf log $\rm <tag>$} : log part of a session between user-defined
        checkpoints
  \item {\bf extract $\rm <tag>$} : reproduce a virtual data product.
\end{itemize}

These commands are implemented by subclassing basic {\tt ROOT} classes.
Commands not recognized by our system are delegated to {\tt ROOT} for
execution or catching exceptions.

The {\tt ROOT} framework provides the ability to access both local and
remote files and datasets. One way to realize the second option is to store
datasets on {\tt APACHE}~\cite{apache} servers
which are {\tt ROOT}-enabled with a
plug-in provided by the {\tt ROOT} team. In this way we implement a
remote data service from Web servers.

The {\tt CVS} system also is able to access both local and remote repositories.
One way to realize the second option is to use the {\tt CVS} pserver.
Contrary to some opinions it can be configured in quite a secure and efficient
way as follows: the remote users need just {\tt CVS} accounts with password
authentication, they never get UNIX accounts on the server. A dedicated
{\tt CVS} user (or several for different groups) acts on their behalf on the
server. The mapping is done by a special {\tt CVS} administrative file.
Similar design was adopted by the {\tt Globus}~\cite{globus} toolkit
with the grid
mapfiles to control user access to remote sites, the only difference
being the use of certificates in place of passwords, thus enhancing the
security and providing a temporarily limited single sign-on to a Grid.
The virtual organization tools developed by {\tt Globus} can be used also
for {\tt CVS} services. In addition, we implement access control lists
per {\tt CVS} user for reading of the repository or writing to specific
directories only for authorized users. This makes the server secure:
even compromising the password a normal user can not modify administrative
files and thus can not run shell commands on the server. Only the
administrator needs an UNIX server account. Adding and managing {\tt CVS} users
is undoubtedly much simpler, safer and more scalable compared to dealing with
UNIX accounts, not to talk about the dreaded group variety. A single server can
handle multiple repositories, making it easy to ``fine structure'' projects.

To test the functionality after each modification we have a test suite:
we use events generated with {\tt PYTHIA}~\cite{pitia} (or the CMS {\tt PYTHIA}
implementation in {\tt CMKIN}), and analyze, histogram and visualize them with
code~\cite{pythview} (for results obtained
with this code see e.g.~\cite{Bourilkov:2003kj,Bourilkov:2003kk})
built on top of the object-oriented data analysis framework
{\tt ROOT}~\cite{root}. To conclude we give a couple of 
snapshots of the first {\tt CAVES} system in action:

{\footnotesize
\begin{verbatim}

START a virtual data enabled ROOT client: rltest

rltest

*************************************************************************
*                        Welcome To CAVES                               *
*       Collaborative Analysis Versioning Environment System            *
*                                                                       *
*                  Dimitri Bourilkov & Mandar Kulkarni                  *
*                       University of Florida                           *
*                         Gainesville, USA                              *
*                                                                       *
*                  You are Welcome to visit our website                 *
*                       cern.ch/bourilkov/caves.html                    *
*************************************************************************

 Please set the cvs pserver OR hit enter for default 

CAVES: 

 Pserver for this session:  :pserver:test@ufgrid02.phys.ufl.edu:/home/caves 

*************************************************************************
*                        TO GET STARTED:                                *
*                  just type  help  at the command prompt               *
*                                                                       *
*            commands beginning with '.' are delegated to ROOT          *
*************************************************************************
CAVES: help
 
List of commands and how to use them: 
 
  === help : to get this help 
 
    help 
 
  === browse : to list all existing tags or a subset beginning with string
  ===>       de facto the content of the virtual data catalog is displayed
 
    browse
    browse <prefix-string-of-tag>
 
  === log : to store all command line activities of the user labeling them with a tag
  ===         the actions after the last log command OR
  ===         from the beginning of the session are stored
  ===         
  ===         <tag> must be CVS compliant i.e. start with uppercase or lowercase letter
  ===         and contain uppercase and lowercase letters, digits, '-' and '_' 
  ===         HINT: this a powerful tool to structure your project 
  ===         
  ===>      in effect this logs how a chunk of virtual data was produced
  ===>      and can be (re)produced later; the macro files executed by the user
  ===>      are stored in their entirety along with the command line activities
  ===>      creating a complete log
 
    log <tag>
 
  === extract : to produce a chunk of virtual data identified by a tag
  ===>      the necessary macro files are downloaded automatically to the client
  ===>      and can be reused and modified for new analyses
 
    extract <tag>
 
CAVES: browse
higgs-ww-plotpxpypz-500  	(revision: 1.5)

higgs-ww-plotpxpypz-100  	(revision: 1.4)



CAVES: extract higgs-ww-plotpxpypz-500
********************Storing data for usage....*********************
ROOT Command is :.x 
Macro is :dbpit1web.C 
Macro is :dbpit1web.C 
U data/dbpit1web.C
You have [0] altered files in this repository.
Are you sure you want to release (and delete) directory `data': y
Argument is 500 
Argument is input "http://ufgrid02.phys.ufl.edu/~bourilkov/higgs.root" 
Argument is output "higgs-ww-plotpxpypz-500" 
Command is :.x dbpit1web.C(500,"http://ufgrid02.phys.ufl.edu/~bourilkov/higgs.root",
"higgs-ww-plotpxpypz-500")
TFile**		higgs-ww-plotpxpypz-500.root	
 TFile*		higgs-ww-plotpxpypz-500.root	
  KEY: TCanvas	canv2;1	ROOT PYTHIA Plotter  D.Bourilkov  University of Florida
.x dbpit1web.C(500,"http://ufgrid02.phys.ufl.edu/~bourilkov/higgs.root",
"higgs-ww-plotpxpypz-500")

**************************End****************************
You have [0] altered files in this repository.
Are you sure you want to release (and delete) directory `v01': y
CAVES: .q
.q
\end{verbatim}
}

\begin{figure*}[htb]
\centering
\resizebox{0.855\textwidth}{0.45\textheight}{\includegraphics{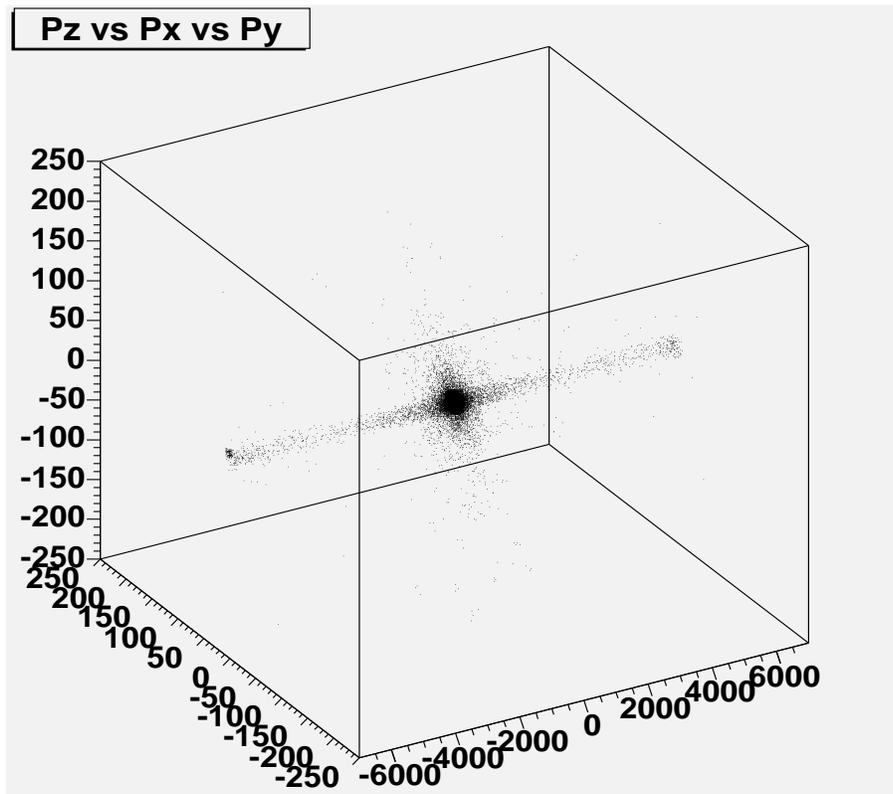}}
\caption{Example of a plot produced with CAVES: 3D momentum
distribution of final state particles for events where the elusive Higgs boson
is produced and decays to a pair of W bosons usually at high transverse
momentum. The LHC beam line and the ``blob'' from the Higgs decays are
clearly visible.}
\label{higgsplot}
\end{figure*}

The running of this example produces the following plot from five hundred
simulated input events, as illustrated in Figure~\ref{higgsplot}.
\begin{figure*}[htb]
\centering
\resizebox{0.855\textwidth}{0.45\textheight}{\includegraphics{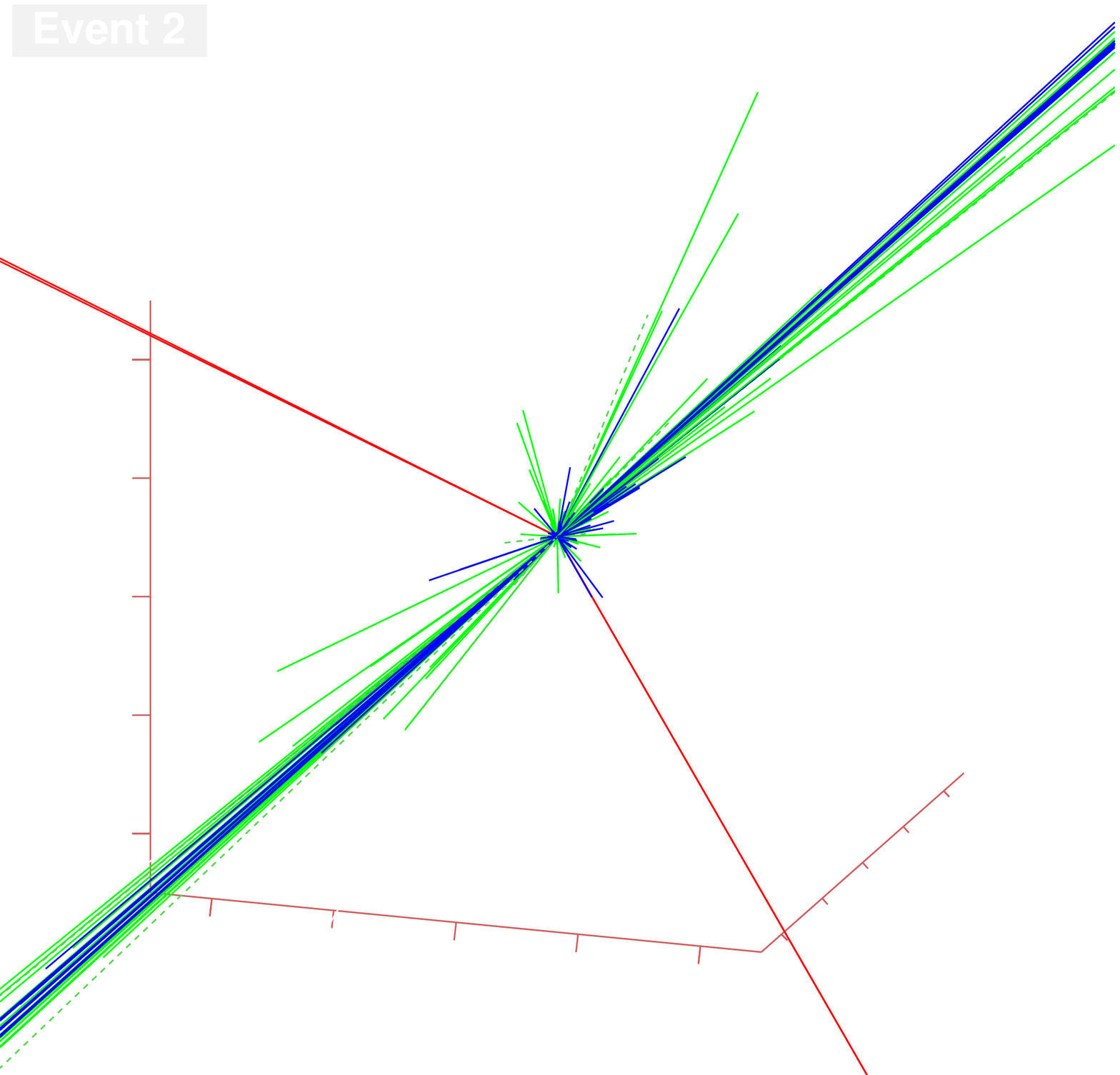}}
\caption{Example of an event display produced with CAVES: production of a
Z boson decaying to muons (the isolated red lines) at high transverse momentum.}
\label{eventdisplay}
\end{figure*}

It is worth mentioning that the plot materializes on the client machine out of
``thin air''. The user can first download the {\tt CAVES} code from our remote
repository and needs just {\tt ROOT} and {\tt CVS} to build the client in no
time. Then the remote logs are browsed, one is selected for extraction,
the corresponding commands and programs are downloaded, built on the fly
and executed on the client machine, the input data are accessed from our
remote Web data server, ``et voil\`{a}'', the plot pops up on the client's
machine.

An example of our event display built also on top of {\tt ROOT} is
shown in Figure~\ref{eventdisplay}.

\section{RELATIONSHIP WITH OTHER GRID PROJECTS}

We envisage extending the service oriented architecture by 
encompassing Grid and Web services as they mature and provide the performance
needed to meet mission-critical requirements.
Our system will benefit from developments like 
the {\tt Globus} authentication system, enhancing the
security and providing a temporarily limited single sign-on to Grid services,
the GriPhyN Virtual Data Toolkit for job executions on Grids, possibly
augmented by Grid schedulers like {\tt SPHINX} or {\tt PEGASUS}.
Other promising developments are remote data services e.g. {\tt CLARENS},
which can possibly be used also with our set of {\tt CVS} services
to provide a Globus Security Infrastructure,
distributed databases etc.
Another closely watched development is the GriPhyN virtual data system
{\tt Chimera}, which evolves a virtual data language. Our ``virtual data
logbooks'' in the first implementation are formatted as standard ASCII files,
in future versions we might use also a more structured format e.g. XML,
which is well suited for Web services.
{\tt Chimera} also converts the virtual data transformations and derivations
to XML and further to directed acyclic graphs. It is a challenging
research question if all analysis activities can be expressed easily
in the present {\tt Chimera} language. With the developments of both projects it
may be possible to generate automatically virtual data language derivations
from our logs of interactive {\tt ROOT}  sessions
for execution on a Grid or storage in a {\tt Chimera} virtual data catalog.

Another promising path is integration with the
{\tt ROOT}/{\tt PROOF}~\cite{Ballintijn:2003yt}
system for parallel analysis of large data sets.

{\tt CAVES} can be used as a building block for a collaborative analysis
environment in a Web and Grid services oriented architecture, important at
a time when Web and Grid services are gaining in prominence.
We are monitoring closely the evolving architecture and use cases of projects
like CAIGEE~\cite{caigee}, HEPCAL~\cite{hepcal}, ARDA~\cite{arda} and
collaborative workflows~\cite{workflow},
and are starting useful collaborations.
Besides the {\tt ROOT}-based client a Web browser executing commands on a
remote {\tt ROOT} or {\tt CLARENS} server is a possible development for
``ultralight'' clients.

\section{OUTLOOK}

In this white paper we have developed the main ideas driving the {\tt CAVES}
project for exploring virtual data concepts for data analysis. The
decomposition of typical analysis tasks shows that the virtual data
approach bears great promise for qualitatively enhancing the collaborative
work of research groups and the accumulation and sharing of knowledge in
todays complex and large scale scientific environments. The confidence in
results and their discovery and reuse grows with the ability to automatically
log and reproduce them on demand.

We have built a first functional system providing automatic
data provenance in a typical analysis session. The system has been demonstrated
successfully at Supercomputing 2003 and a first public release is available
for interested users, which are encouraged to visit our Web pages, currently
located at the following URL:\\
{\tt http://cern.ch/bourilkov/caves.html }

\begin{acknowledgments}
In the process of shaping and launching the project the author enjoyed
discussions with Paul Avery, Rene Brun, Federico Carminati, Harwey Newman,
Lothar Bauerdick, David Stickland, Stephan Wynhoff, Torre Wenaus, Rob Gardner,
Fons Rademakers, Richard Cavanaugh, John Yelton, Darin Acosta, Mike Wilde,
Jens V\"{o}ckler, Conrad Steenberg, Julian Bunn, Predrag Buncic
and Martin Ballantijn.
The coffee pauses with Jorge Rodriguez were always stimulating.
The first prototype and first release are coded by Mandar Kulkarni and myself,
the members of the {\tt CAVES} team at present.
This work is supported in part by the United States National Science Foundation
under grants NSF ITR-0086044 (GriPhyN) and NSF PHY-0122557 (iVDGL).

\end{acknowledgments}


\end{document}